\documentclass[prd,preprint,superscriptaddress,showpacs,byrevtex]{revtex4}
\usepackage{bm}
\def\fsl#1{\setbox0=\hbox{$#1$}           % set a box for #1
   \dimen0=\wd0                                 % and get its size
   \setbox1=\hbox{/} \dimen1=\wd1               % get size of /
   \ifdim\dimen0>\dimen1                        % #1 is bigger
      \rlap{\hbox to \dimen0{\hfil/\hfil}}      % so center / in box
      #1                                        % and print #1
   \else                                        % / is bigger
      \rlap{\hbox to \dimen1{\hfil$#1$\hfil}}   % so center #1
      /                                         % and print /
   \fi}                                         %
\newcommand{\be}{\begin{equation}}
\newcommand{\ee}{\end{equation}}
\newcommand{\bea}{\begin{eqnarray}}
\newcommand{\eea}{\end{eqnarray}}
\newcommand{\beq}{\begin{equation}}
\newcommand{\eeq}{\end{equation}}
\newcommand{\beqs}{\begin{eqnarray}}
\newcommand{\eeqs}{\end{eqnarray}}

\newcommand{\aslash}{A\hspace{-0.067in}\slash}

\begin{document}
\title{ Lattice QCD Method To Study Proton Decay }
\author{Gouranga C Nayak }\thanks{E-Mail: nayakg138@gmail.com}
%
%\affiliation{ C. N. Yang Institute for Theoretical Physics, Stony Brook University, Stony Brook NY, 11794-3840 USA}
%
\date{\today}
\begin{abstract}
The proton decay has not been experimentally observed with the lower limit of the proton lifetime being $\gtrsim 10^{34}$ years which is more than the age of the universe. One of the important quantity that appears in the study of the proton decay is the proton decay matrix element which is a non-perturbative quantity in QCD which cannot be calculated by using the perturbative QCD (pQCD) method but it can be calculated by using the lattice QCD method. In this paper we formulate the lattice QCD method to study the proton decay matrix element. We derive the non-perturbative formula of the proton decay matrix element from the first principle in QCD which can be calculated by using the lattice QCD method.
\end{abstract}
\maketitle
\pagestyle{plain}

\pagenumbering{arabic}

\section{ Introduction }

The decay of the proton has not been experimentally observed since the universe was created almost 14 billion years ago. On the other hand the other hadrons such as neutron, pion and kaon etc. decay with finite lifetime. For example, the lifetime of the neutron is $\sim$ 880 seconds, the lifetime of neutral pion is $\sim 10^{-16}$ seconds and the lifetime of charged kaon is $\sim 10^{-8}$ seconds. In comparison to this the lower limit of the proton lifetime is $\gtrsim 10^{34}$ years which is more than the age of the universe.

The main reason why the proton decay has not been observed is due to the baryon number conservation. Note that in the standard model of  physics the baryon number is conserved. For example, the baryon number is conserved in the (free) neutron decay $n ~\rightarrow ~p +e +{\bar \nu}_e$
where $n$ is the neutron, $p$ is the proton, $e$ is the electron and ${\bar \nu}_e$ is the electron type antineutrino. However, the (free) proton decay processes such as
\bea
p \rightarrow \pi_0 + e^+,~~~~~~~~~~~~~p \rightarrow \pi_0 + \mu^+
\label{ppe}
\eea
are not allowed because the baryon number is not conserved for the processes in eq. (\ref{ppe}) in the standard model of physics where $\pi_0$ is the neutral pion, $e^+$ is the positron and $\mu^+$ is the muon. It is well known that the (free) proton decay process $p ~ \rightarrow~ n +e^+ +\nu_e$ is not allowed even if the baryon number is conserved because the neutron mass is larger than the proton mass where $\nu_e$ is the electron type neutrino. In this paper we refer the (free) proton decay as the proton decay.

In the beyond standard model of physics the baryon number violation can occur which can lead to the proton decay. For example, the beyond standard model of  physics such as the grand unified theories (GUTs) and the supersymmetry grand unified theories (SUSY-GUTs) predict the proton decay.

Note that even if the beyond standard model of physics predicts the proton decay but as mentioned above the proton decay has not been experimentally observed since the universe was created almost 14 billion years ago. Over the several decades various experiments have searched for the proton decay although these experiments have not found any clear evidence of the proton decay. By comparing these experimental searches with the parameter spaces of the GUTs and
SUSY-GUTs these experimental searches have imposed tight constraints into the parameter spaces of the GUTs and SUSY-GUTs.

For the proton decay channels $p \rightarrow \pi_0 + e^+$ and $p \rightarrow \pi_0 + \mu^+$ in eq. (\ref{ppe})
the Super-Kamiokande experiment \cite{sk} has imposed the lower limit of the proton decay lifetime to be
$t_{p \rightarrow \pi_0 + e^+} > 8.2 \times 10^{33}$ years and $t_{p \rightarrow \pi_0 + \mu^+} > 6.6 \times 10^{33}$ years respectively.
For the proton decay channel
\bea
p \rightarrow K^+ +  \nu
\label{pkn}
\eea
the Super-Kamiokande experiment \cite{sk1} has imposed the lower limit of the proton decay lifetime to be $t_{p \rightarrow K^+ + \nu } > 5.9 \times 10^{33}$ years where $K^+$ is the positively charged kaon.

The initial state for the proton decay channels in eq. (\ref{ppe}) is $|p>$ and the final states are $|\pi_0 e^+>$ and $|\pi_0 \mu^+>$ respectively. Since the leptons $e^+$ and $\mu^+$ in the final states can be treated trivially one needs to calculate the matrix element $<\pi_0|{\cal O}^{\not B}|p>$ to study the proton decay where ${\cal O}^{\not B}$ is the three-quark operator violating the baryon number [see eq. (\ref{mea})].

The proton and pion consist of quarks, antiquarks and gluons which are described by the quantum chromodynamics (QCD) \cite{ymk} which is a fundamental theory of the nature. The partonic cross section at the short distance can be calculated by using the perturbative QCD (pQCD) due to asymptotic freedom in QCD \cite{gwk}. The factorization theorem in QCD \cite{fck,fck1,fck2} plays a central role to calculate the hadron cross section from the parton cross section at the high energy colliders.

The hadron formation from the quarks and gluons is a long distance phenomena in QCD which cannot be studied by using the pQCD but can be studied by using the non-perturbative QCD. Hence the proton decay matrix element $<\pi_0|{\cal O}^{\not B}|p>$ is a non-perturbative matrix element in QCD which cannot be calculated by using perturbative QCD but can be calculated by using the non-perturbative QCD. On the other hand the analytical solution of the non-perturbative QCD is not known yet. Hence the lattice QCD method can be used to calculate the proton decay matrix element $<\pi_0|{\cal O}^{\not B}|p>$.

Recently we have presented the lattice QCD method to study the proton formation from the quarks and gluons \cite{pqg} and to study the proton spin crisis \cite{psc} by implementing the non-zero boundary surface term in QCD due to the confinement of quarks and gluons inside the finite size proton \cite{nkbs}.

In this paper we extend this to study the proton decay matrix element and present the lattice QCD formulation to study the proton decay matrix element by implementing this non-zero boundary surface term in QCD due to confinement. We derive the non-perturbative formula of the proton decay matrix element $<\pi_0|{\cal O}^{\not B}|p>$  from the first principle in QCD at all orders in coupling constant which can be calculated by using the lattice QCD method by implementing this non-zero boundary surface term in QCD due to confinement. Extension of this procedure to calculate the other proton decay matrix elements such as $<K^+|{\cal O}^{\not B}|p>$ is straightforward.

The paper is organized as follows. In section II we describe the lattice QCD method to study the proton formation from quarks and gluons by implementing the non-zero boundary surface term in QCD due to confinement. In section III we present the formulation of the lattice QCD method to study the proton decay matrix element by implementing this non-zero boundary surface term in QCD due to confinement. Section IV contains conclusions.

\section{ Proton formation from quarks and gluons using lattice QCD Method }

We denote the up and down quark fields by $u_i(x)$ and $d_i(x)$ respectively where $i=1,2,3$ is the color index. The partonic operator to study the proton formation from the partons is given by
\bea
{\cal O}_p(x)=\epsilon_{ijk} u^T_i(x) C \gamma_5 d_j(x) u_k(x)
\label{po}
\eea
where $C$ is the charge conjugation operator. The time evolution of the partonic operator ${\cal O}_p(x)$ is given by
\bea
{\cal O}_p(r,t) = e^{-iHt} {\cal O}_p(r,0) e^{iHt}
\label{te}
\eea
where $H$ is the QCD hamiltonian of the partons.

The vacuum expectation value of the two point correlation function of the partonic operators in QCD is given by
\bea
&& <0|{\cal O}_p(r',t') {\cal O}_p(0)|0> = \frac{1}{Z[0]} \int [d{\bar u}][du] [d{\bar d}][dd]~{\cal O}_p(r',t') {\cal O}_p(0)~ {\rm det}[\frac{\delta G_f^c}{\delta \omega^a}] \nonumber \\
&&\times {\rm exp}[i\int d^4x [-\frac{1}{4} F_{\nu \lambda}^d(x) F^{\nu \lambda d}(x) -\frac{1}{2\alpha} [G_f^c(x)]^2 +{\bar u}_k(x)[\delta^{kj}(i{\not \partial}-m_u) +gT^d_{kj}\aslash^d(x)]u_j(x) \nonumber \\
&&+{\bar d}_k(x)[\delta^{kj}(i{\not \partial}-m_d) +gT^d_{kj}\aslash^d(x)]d_j(x)]]
\label{cfa}
\eea
where $|0>$ is the vacuum state of the full QCD (not pQCD), $A_\mu^a(x)$ is the gluon field, $G_f^c(x)$ is the gauge fixing term, $\alpha$ is the gauge fixing parameter and
\bea
&& Z[0]= \int [d{\bar u}][du] [d{\bar d}][dd]~ {\rm det}[\frac{\delta G_f^c}{\delta \omega^a}] \times {\rm exp}[i\int d^4x [-\frac{1}{4} F_{\nu \lambda}^d(x) F^{\nu \lambda d}(x) -\frac{1}{2\alpha} [G_f^c(x)]^2 \nonumber \\
&&+{\bar u}_k(x)[\delta^{kj}(i{\not \partial}-m_u) +gT^d_{kj}\aslash^d(x)]u_j(x) +{\bar d}_k(x)[\delta^{kj}(i{\not \partial}-m_d) +gT^d_{kj}\aslash^d(x)]d_j(x)]]
\label{cfb}
\eea
is the generating functional in QCD with
\bea
F_{\nu \lambda}^s(x) = \partial_\nu A_\lambda^s(x) - \partial_\lambda A_\nu^s(x)+gf^{sab} A_\nu^a(x) A_\lambda^b(x).
\label{cfc}
\eea
Note that the ghost fields are absent in eq. (\ref{cfa}) because we are directly dealing with the ghost determinant ${\rm det}[\frac{\delta G_f^c}{\delta \omega^a}]$ in this paper.

The complete set of hadronic energy-momentum eigenstates is given by
\bea
\sum_{n'} |H_{n'}><H_{n'}|=1.
\label{cfd}
\eea
Using eqs. (\ref{te}) and (\ref{cfd}) in (\ref{cfa}) we find in the Euclidean time
\bea
&& \sum_r <0|{\cal O}^\dagger_p(r,t) {\cal O}_p(0)|0> =\sum_{n'}|<H_{n'}|{\cal O}_p(0)|0>|^2e^{-\int dt E_{n'}(t)}
\label{cfe}
\eea
where $\int dt$ is an indefinite integration and $E_n(t)$ is the energy of all the partons inside the proton in its $n$th energy level state which is time dependent [see eq. (\ref{cfn})] given by
\bea
H|H_{n'}>=E_{n'}(t)|H_{n'}>.
\label{cff}
\eea
Neglecting the higher energy level contributions at the large time we find
\bea
&& [\sum_r  <0|{\cal O}^\dagger_p(r,t) {\cal O}_p(0)|0>]_{t \rightarrow \infty} =|<p|{\cal O}_p(0)|0>|^2e^{-\int dt E_p(t)}
\label{cfg}
\eea
where $|p>$ is the energy-momentum eigenstate of the proton $p$, the $E_p(t)$ is the energy of all the partons inside the proton given by
\bea
H|H_{0}>=E_{0}(t)|H_{0}>,~~~~~~~|p>=|H_0>,~~~~~~~~~~~E_p(t)=E_0(t).
\label{cfh}
\eea
In terms of the energy-momentum tensor of the partons inside the proton we find
\bea
E_p(t)=<p|\sum_{q,{\bar q},g} \int d^3r T^{00}(r,t)|p>
\label{cfi}
\eea
where $T^{\mu \nu}(x)$ is the energy-momentum tensor density in QCD given by
\bea
&&T^{\nu \lambda}(x) = F^{\nu \mu d}(x) F_{\mu}^{~\lambda d}(x) +\frac{1}{4} g^{\nu \lambda} F_{\mu \sigma}^d(x)F^{\mu \sigma d}(x)
+ {\bar u}_k(x) \gamma^\nu [i\partial^\lambda -igT^d_{kj}A^{\lambda d}(x)]u_j(x)\nonumber \\
&& + {\bar d}_k(x) \gamma^\nu [i\partial^\lambda -igT^d_{kj}A^{\lambda d}(x)]d_j(x).
\label{cfj}
\eea
From the continuity equation $\partial_\nu T^{\nu \lambda}(x)=0$ we obtain
\bea
\frac{d}{dt}<p|\int d^3r T^{00}(r,t)|p>=-<p|\int d^3r \partial_k T^{k0}(r,t)|p>.
\label{cfl}
\eea
Due to the confinement of quarks and gluons inside the finite size proton we find the non-zero boundary surface term in QCD \cite{nkbs}
\bea
<p|\int d^3r \partial_k T^{k0}(r,t)|p>\neq 0
\label{cfm}
\eea
which from eqs. (\ref{cfi})  and (\ref{cfl}) gives
\bea
\frac{dE_p(t)}{dt}\neq 0.
\label{cfn}
\eea
Hence from eq. (\ref{cfn}) we find that the energy $E_p(t)$ of all the quarks, antiquarks and gluons inside the proton is not constant but is time dependent. Since the energy $E_p$ of the proton is constant (time independent) we find that
\bea
E_p\neq E_p(t)
\label{cfo}
\eea
where $E_p(t)$ is the energy of all the partons inside the protopn $p$ and $E_p$ is the energy of the proton $p$. From eqs. (\ref{cfi}), (\ref{cfl}) and (\ref{cfm}) we obtain
\bea
\frac{d}{dt}[E_p(t)+E_B(t)]=0
\label{cfp}
\eea
where
\bea
\frac{dE_B(t)}{dt}=<p|\sum_{q,{\bar q},g} \int d^3r \partial_k T^{k0}(r,t)|p>\neq 0
\label{cfq}
\eea
Hence, unlike eqs. (\ref{cfn}) and (\ref{cfo}), we find from eq. (\ref{cfp}) that
\bea
E_p=E_p(t)+E_B(t)
\label{cfr}
\eea
where $E_p(t)$ is given by eq. (\ref{cfi}) and $E_B(t)$ is given by eq. (\ref{cfq}).

Using eq. (\ref{cfr}) in (\ref{cfg}) we find
\bea
&& |<p|{\cal O}_p(0)|0>|^2e^{- m_pt}=[\frac{\sum_r <0|{\cal O}^\dagger_p(r,t) {\cal O}_p(0)|0>}{e^{\int dt E_B(t)}}]_{t \rightarrow \infty}
\label{cfs}
\eea
where $m_p$ is the mass of the proton.

The vacuum expectation value of the three point correlation function of the partonic operators in QCD is given by
\bea
&& <0|{\cal O}_p(r'',t'') [\int d^3r' \partial_k T^{k0}(r',t')] {\cal O}_p(0)|0> = \frac{1}{Z[0]} \int [d{\bar u}][du] [d{\bar d}][dd]~{\cal O}_p(r'',t'') \nonumber \\
&&\times [\int d^3r' \partial_k T^{k0}(r',t')]\times {\cal O}_p(0)~ {\rm det}[\frac{\delta G_f^c}{\delta \omega^a}] \times {\rm exp}[i\int d^4x [-\frac{1}{4} F_{\nu \lambda}^d(x) F^{\nu \lambda d}(x) -\frac{1}{2\alpha} [G_f^c(x)]^2 \nonumber \\
&&+{\bar u}_k(x)[\delta^{kj}(i{\not \partial}-m_u) +gT^d_{kj}\aslash^d(x)]u_j(x) +{\bar d}_k(x)[\delta^{kj}(i{\not \partial}-m_d) +gT^d_{kj}\aslash^d(x)]d_j(x)]]. \nonumber \\
\label{cft}
\eea
Using eqs. (\ref{te}) and (\ref{cfd}) in (\ref{cft}) we find in the Euclidean time
\bea
&& \sum_{r'} <0|{\cal O}^\dagger_p(r',t')[\sum_{q,{\bar q},g}\int d^3r \partial_k T^{k0}(r,t)] {\cal O}_p(0)|0> =\sum_{n',n''}<0|{\cal O}^\dagger_p(0)|H_{n'}>\nonumber \\
&&<H_{n'}|[\sum_{q,{\bar q},g}\int d^3r \partial_k T^{k0}(r,t)]|H_{n''}><H_{n''}|{\cal O}_p(0)|0>|^2e^{-\int dt' E_{n'}(t')}.
\label{cfu}
\eea
Neglecting the higher energy level contributions at the large time we obtain
\bea
&& [\sum_{r'} <0|{\cal O}^\dagger_p(r',t')[\sum_{q,{\bar q},g}\int d^3r \partial_k T^{k0}(r,t)] {\cal O}_p(0)|0>]_{t' \rightarrow \infty} \nonumber \\
&& =<p|\sum_{q,{\bar q},g}\int d^3r \partial_k T^{k0}(r,t)|p>|<p|{\cal O}_p(0)|0>|^2e^{-\int dt' E(t')}.
\label{cfv}
\eea
From eqs. (\ref{cfg}), (\ref{cfv}) and (\ref{cfq}) we find
\bea
\frac{dE_B(t)}{dt}=[\frac{\sum_{r'} <0|{\cal O}^\dagger_p(r',t')[\sum_{q,{\bar q},g}\int d^3r \partial_k T^{k0}(r,t)] {\cal O}_p(0)|0>}{\sum_{r'} <0|{\cal O}^\dagger_p(r',t'){\cal O}_p(0)|0>}]_{t' \rightarrow \infty}.
\label{cfw}
\eea
Using eq. (\ref{cfw}) in (\ref{cfs}) we obtain
\bea
&& |<p|{\cal O}_p(0)|0>|^2e^{- m_pt}=[\frac{\sum_r <0|{\cal O}^\dagger_p(r,t) {\cal O}_p(0)|0>}{e^{\int dt [\frac{\sum_{r'} <0|{\cal O}^\dagger_p(r',t')[\sum_{q,{\bar q},g}\int dt \int d^3r \partial_k T^{k0}(r,t) ]{\cal O}_p(0)|0>}{\sum_{r'} <0|{\cal O}^\dagger_p(r',t'){\cal O}_p(0)|0>}]_{t' \rightarrow \infty}}}]_{t \rightarrow \infty}\nonumber \\
\label{cfx}
\eea
where $\int dt$ is indefinite integration.

Eq. (\ref{cfx}) is the non-perturbative formula to study the proton formation from quarks, antiquarks and gluons by implementing the non-zero boundary surface term in QCD due to confinement which can be calculated by using the lattice QCD method.

\section{Lattice QCD Method To Study Proton Decay }

In this section we will extend the procedure of the previous section to derive the non-perturbative formula of the proton decay matrix element $<\pi_0|{\cal O}^{\not B}|p>$ by implementing the non-zero boundary surface term in QCD due to confinement which can be calculated by using the lattice QCD method. This procedure is also applied to study various non-perturbative quantities in QCD in vacuum \cite{psc,allg} and in QCD in medium \cite{allgm} to study the quark-gluon plasma at RHIC and LHC \cite{qgk,qgk1,qgk2}.

The baryon number violating three-quark operator ${\cal O}^{\not B}(x)$ is given by
\bea
{\cal O}^{\not B}(x) =\epsilon^{ijk} [u_i^T(x)CP_{R/L}d_j(x)]P_L u_k(x)
\label{mea}
\eea
where $P_{R/L}$ means right or left projection matrix respectively given by
\bea
P_R=\frac{1+\gamma_5}{2},~~~~~~~P_L=\frac{1-\gamma_5}{2}.
\label{mec}
\eea
The partonic operator ${\cal O}_p(x)$ for the proton formation is given by eq. (\ref{po}) and the patonic operator ${\cal O}_{\pi_0}(x)$ for the pion formation is given by
\bea
{ {\cal O}}_{\pi_0}(x) =\frac{{\bar u}_i(x)\gamma_5 u_i(x)-{\bar d}_i(x)\gamma_5 d_i(x)}{\sqrt{2}}.
\label{med}
\eea
The vacuum expectation value of the three point non-perturbative partonic correlation function $<0|{\cal O}_{\pi_0}(r',t'){\cal O}^{\not B}(r,t) {\cal O}_p(0)|0>$ is given by
\bea
&& <0|{\cal O}_{\pi_0}(r'',t''){\cal O}^{\not B}(r',t') {\cal O}_p(0)|0> = \frac{1}{Z[0]} \int [d{\bar u}][du] [d{\bar d}][dd]~{\cal O}_{\pi_0}(r'',t'') {\cal O}^{\not B}(r',t'){\cal O}_p(0)\nonumber \\
&&\times {\rm det}[\frac{\delta G_f^c}{\delta \omega^a}] \times {\rm exp}[i\int d^4x [-\frac{1}{4} F_{\nu \lambda}^d(x) F^{\nu \lambda d}(x) -\frac{1}{2\alpha} [G_f^c(x)]^2 +{\bar u}_k(x)[\delta^{kj}(i{\not \partial}-m_u) +gT^d_{kj}\aslash^d(x)]u_j(x) \nonumber \\
&&+{\bar d}_k(x)[\delta^{kj}(i{\not \partial}-m_d) +gT^d_{kj}\aslash^d(x)]d_j(x)]].
\label{mee}
\eea
Similar to eq. (\ref{cfd}) for the proton the complete set of energy-momentum eigenstates of the pion $\pi_0$ is given by
\bea
\sum_{n'} |H^{\pi_0}_{n'}><H^{\pi_0}_{n'}|=1.
\label{cfdpi}
\eea
Using eqs. (\ref{te}), (\ref{cfd}) and (\ref{cfdpi}) in (\ref{mee}) we find in the Euclidean time
\bea
&& \sum_{r'',r'} e^{i{\vec p}_{\pi_0}\cdot ({\vec r}''-{\vec r}')} <0|{\cal O}_{\pi_0}(r'',t''){\cal O}^{\not B}(r',t') {\cal O}_p(0)|0> =\sum_{n'',n'}<0|{\cal O}_{\pi_0}|H^{\pi_0}_{n''}>\nonumber \\
&&<H^{\pi_0}_{n''}|{\cal O}^{\not B}|H_{n'}><H_{n'}|{\cal O}_p|0>e^{-[\int dt'' E^{\pi_0}_{n''}(t'')-\int dt' E^{\pi_0}_{n''}(t')]}e^{-\int dt' E^{p}_{n'}(t')}
\label{mef}
\eea
where $\int dt'$ and $\int dt''$ are indefinite integrations, ${\vec p}_{\pi_0}$ is the momentum of the pion $\pi_0$ and the proton is at rest.

In the limit $t'' >>>t',~~t'\rightarrow \infty$ we find by neglecting the higher energy level contributions
\bea
&& [\sum_{r'',r'} e^{i{\vec p}_{\pi_0}\cdot ({\vec r}''-{\vec r}')} <0|{\cal O}_{\pi_0}(r'',t''){\cal O}^{\not B}(r',t') {\cal O}_p(0)|0>]_{t''>>>t',~~t\rightarrow \infty} =<0|{\cal O}_{\pi_0}|\pi({\vec p}_{\pi_0})>\nonumber \\
&&<\pi({\vec p}_{\pi_0})|{\cal O}^{\not B}|p><p|{\cal O}_p|0>e^{-[\int dt'' E_{\pi_0}(t'')-\int dt' E_{\pi_0}(t')]}e^{-\int dt' E_{p}(t')}
\label{meg}
\eea
where $E_{\pi_0}(t)$ is the energy of all the partons inside the pion $\pi_0$ and $E_p(t)$ is the energy of all the partons inside the proton $p$.

From eq. (\ref{cfg}) we find for the proton formation
\bea
&& [\sum_{r'} <0|{\cal O}_p(r',t') {\cal O}_p(0)|0>]_{t' \rightarrow \infty} =|<p|{\cal O}_p|0>|^2e^{-\int dt' E_{p}(t')}.
\label{meh}
\eea
Similarly for the pion $\pi_0$ formation we find
\bea
&& [\sum_{r''} e^{i{\vec p}_{\pi_0}\cdot {\vec r}''} <0|{\cal O}_{\pi_0}(r'',t''-t') {\cal O}_{\pi_0}(0)|0>]_{t''>>>t',~~t' \rightarrow \infty} \nonumber \\
&&=|<\pi_0({\vec p}_{\pi_0})|{\cal O}_{\pi_0}|0>|^2e^{-[\int dt'' E_{\pi_0}(t'')-\int dt' E_{\pi_0}(t')]}.
\label{mei}
\eea
From eqs. (\ref{meg}), (\ref{meh}) and (\ref{mei}) we find
\bea
&&<\pi_0({\vec p}_{\pi_0})|{\cal O}^{\not B}|p>= \sqrt{|<\pi({\vec p}_{\pi_0})|{\cal O}_{\pi_0}|0>|^2}\times \sqrt{|<p|{\cal O}_p|0>|^2}\nonumber \\
&&\times [\frac{\sum_{r',r''} e^{i{\vec p}_{\pi_0}\cdot ({\vec r}''-{\vec r}')} <0|{\cal O}_{\pi_0}(r'',t''){\cal O}^{\not B}(r',t') {\cal O}_p(0)|0>}{[\sum_{r'} <0|{\cal O}_p(r',t') {\cal O}_p(0)|0>][\sum_{r''} e^{i{\vec p}_{\pi_0}\cdot {\vec r}''} <0|{\cal O}_{\pi_0}(r'',t''-t') {\cal O}_{\pi_0}(0)|0>]}]_{t''>>>t',~~~t'\rightarrow \infty}.\nonumber \\
\label{mej}
\eea
From eq. (\ref{cfx}) we find
\bea
&& |<p|{\cal O}_p(0)|0>|^2=[\frac{\sum_r <0|{\cal O}^\dagger_p(r,t) {\cal O}_p(0)|0>\times e^{ m_pt}}{e^{\int dt [\frac{\sum_{r'} <0|{\cal O}^\dagger_p(r',t')[\sum_{q,{\bar q},g}\int dt \int d^3r \partial_k T^{k0}(r,t)] {\cal O}_p(0)|0>}{\sum_{r'} <0|{\cal O}^\dagger_p(r',t'){\cal O}_p(0)|0>}]_{t' \rightarrow \infty}}}]_{t \rightarrow \infty}\nonumber \\
\label{mek}
\eea
and similarly for the pion we find
\bea
&& |<\pi_0({\vec p}_{\pi_0})|{\cal O}_{\pi_0}(0)|0>|^2=[\frac{\sum_r e^{i{\vec p}_{\pi_0}\cdot {\vec r}}<0|{\cal O}^\dagger_{\pi_0}(r,t) {\cal O}_{\pi_0}(0)|0>\times e^{t E_{\pi_0}}}{e^{\int dt [\frac{\sum_{r'} e^{i{\vec p}_{\pi_0}\cdot {\vec r}'}<0|{\cal O}^\dagger_{\pi_0}(r',t')[\sum_{q,{\bar q},g}\int dt \int d^3r \partial_k T^{k0}(r,t)] {\cal O}_{\pi_0}(0)|0>}{\sum_{r'} e^{i{\vec p}_{\pi_0}\cdot {\vec r}'}<0|{\cal O}^\dagger_{\pi_0}(r',t'){\cal O}_{\pi_0}(0)|0>}]_{t' \rightarrow \infty}}}]_{t \rightarrow \infty}\nonumber \\
\label{mel}
\eea
where $E_{\pi_0}$ is the energy of the pion $\pi_0$.

Using eqs. (\ref{mek}) and (\ref{mel}) in (\ref{mej}) we find
\bea
&&<\pi({\vec p}_{\pi_0})|{\cal O}^{\not B}|p>= \left[[\frac{\sum_r e^{i{\vec p}_{\pi_0}\cdot {\vec r}}<0|{\cal O}^\dagger_{\pi_0}(r,t) {\cal O}_{\pi_0}(0)|0>\times e^{t E_{\pi_0}}}{e^{\int dt [\frac{\sum_{r'} e^{i{\vec p}_{\pi_0}\cdot {\vec r}'}<0|{\cal O}^\dagger_{\pi_0}(r',t')[\sum_{q,{\bar q},g}\int dt \int d^3r \partial_k T^{k0}(r,t)] {\cal O}_{\pi_0}(0)|0>}{\sum_{r'} e^{i{\vec p}_{\pi_0}\cdot {\vec r}'}<0|{\cal O}^\dagger_{\pi_0}(r',t'){\cal O}_{\pi_0}(0)|0>}]_{t' \rightarrow \infty}}}]_{t \rightarrow \infty}\right]^{\frac{1}{2}} \nonumber \\
&& \times \left[[\frac{\sum_r <0|{\cal O}^\dagger_p(r,t) {\cal O}_p(0)|0>\times e^{ m_pt}}{e^{\int dt [\frac{\sum_{r'} <0|{\cal O}^\dagger_p(r',t')[\sum_{q,{\bar q},g}\int dt \int d^3r \partial_k T^{k0}(r,t)] {\cal O}_p(0)|0>}{\sum_{r'} <0|{\cal O}^\dagger_p(r',t'){\cal O}_p(0)|0>}]_{t' \rightarrow \infty}}}]_{t \rightarrow \infty}\right]^{\frac{1}{2}}\nonumber \\
&&\times \left[\frac{\sum_{r',r''} e^{i{\vec p}_{\pi_0}\cdot ({\vec r}''-{\vec r}')} <0|{\cal O}_{\pi_0}(r'',t''){\cal O}^{\not B}(r',t') {\cal O}_p(0)|0>}{[\sum_{r'} <0|{\cal O}_p(r',t') {\cal O}_p(0)|0>][\sum_{r''} e^{i{\vec p}_{\pi_0}\cdot {\vec r}''} <0|{\cal O}_{\pi_0}(r'',t''-t') {\cal O}_{\pi_0}(0)|0>]}\right]_{t''>>>t',~~t'\rightarrow \infty}\nonumber \\
\label{mem}
\eea
which can be calculated by using the lattice QCD method.

Eq. (\ref{mem}) is the non-perturbative formula of the proton decay matrix element $<\pi_0|{\cal O}^{\not B}|p>$ derived from the first principle in QCD which can be calculated by using the lattice QCD method. Extension of eq. (\ref{mem}) to study other proton decay matrix elements such as ~~~~~~~~~~~~~~~~~~$<K^+|{\cal O}^{\not B}|p>$ is straightforward.

\section{Conclusions}
The proton decay has not been experimentally observed with the lower limit of the proton lifetime being $\gtrsim 10^{34}$ years which is more than the age of the universe. One of the important quantity that appears in the study of the proton decay is the proton decay matrix element which is a non-perturbative quantity in QCD which cannot be calculated by using the perturbative QCD (pQCD) method but it can be calculated by using the lattice QCD method. In this paper we have formulated the lattice QCD method to study the proton decay matrix element. We have derived the non-perturbative formula of the proton decay matrix element from the first principle in QCD which can be calculated by using the lattice QCD method.

\end{document}